\documentclass[usenatbib,useAMS]{mn2e}
\usepackage{times}
\usepackage{epsfig}

\renewcommand{\vec}[1]{\bmath{#1}}

\begin{document}
\title{Adaptive Contouring -- an efficient way to calculate microlensing
light curves of extended sources}
\author[M. Dominik]{M. Dominik\thanks{Royal Society University Research Fellow}\thanks{E-mail: md35@st-andrews.ac.uk} \\
SUPA, University of St Andrews, School of Physics \& Astronomy, North Haugh,
St Andrews, KY16 9SS, United Kingdom}

\maketitle

\begin{abstract}
The availability of a robust and efficient routine for calculating light curves of 
a finite source magnified due to bending its light by the gravitational field of an intervening binary 
lens is essential for determining the characteristics of planets in such microlensing events, as
well as for modelling stellar lens binaries and resolving the brightness profile of the source star.
However, the presence of extended caustics and the fact that the images of the source star cannot be
determined analytically while their number depends on the source position (relative to the lens system),
makes such a task difficult in general. Combining the advantages of several earlier approaches, an
adaptive contouring algorithm is presented, which only relies on a small number of simple
rules and operations on the adaptive search grid.
By using the parametric representation of critical curves
and caustics found by \cite{Erdl}, seed solutions to the adaptive grid are found, which ensures that 
no images or holes are missed.

\end{abstract}


\section{Introduction}
In order to determine the characteristics of planets that reveal their existence
in microlensing events, to model stellar lens binaries, and to distinguish between 
these configurations, efficient algorithms for calculating light curves of
extended source stars due to microlensing of a binary lens system are required.
For a single point-like lens, the magnification of a point source is
an analytic function \citep{Ein36}, the magnification of a uniformly bright
circular source can be expressed by means of elliptic integrals \citep{WM94},
and is therefore a one-dimensional integral over a semi-analytic function in general,
which can be approximated by the product of the point-source
magnification with a characteristic function of a single variable, 
depending on the source brightness profile \citep{Gouldapprox}. 
For a binary lens, however, the complexity increases enormously. Whereas a point source affected
by a single point-mass lens always has two images, with the only exception of lens and source being
perfectly aligned, binary lenses yield either three or five images
depending on the source position relative to the lens, where three-image and 
five-image regions are separated by extended caustics. This implies that one cannot as easily
integrate over the source position as in the single-lens case. Moreover, the point-source 
magnification cannot be expressed as analytic or semi-analytic function, but needs to be 
determined by means of solving a fifth-order complex polynomial \citep{WM95:fifth}. While a 
two-dimensional integration over the source is possible in principle, one carefully needs to keep 
track of the caustic and needs to deal with a divergent integrand as it is approached.
While ray-shooting algorithms \citep{KRS:rayshooting,SchneiWei:AGN} overcome the need
for determining the images by creating a magnification map based on mapping the image plane to the 
corresponding true source positions, summing the hits within a source still is a 
discretization of a two-dimensional integration. 
On the other hand, \citet{SK87} have found that contour plot routines offer a simple way
for plotting the images of an extended source, which avoids the inversion of the lens equation. 
\citet{Do95:Num} later showed that in addition to providing plots, the same technique
can be used to efficiently derive numerical values, and in particular to
determine the magnification of extended sources as well as of the incurred
astrometric shift in the centroid of their unresolved imges
by applying Green's theorem \citep{Do98:NumSrc}. In fact, determining
the image area from a contour plot is similar to the ray-shooting 
magnification map approach in mapping an image plane grid to the
respective source position, but rather than the magnification being
determined over the source area, an integral in the image plane is
discretized.

In this paper, an algorithm is presented that determines the magnification
from the image contour using an adaptive grid rather than a static one. 
By recognizing the positions of the images of the source centre, which can
be calculated by solving a fifth-order complex polynomial \citep{WM95:fifth},
the fact that holes in the images must include the
positions of the point-like deflectors, and by finding all images stretching over critical curves using a parametric representation based on the classification
of the topology of the critical curves and caustics by \citet{Erdl},
it is ensured that all images for a binary lens are appropriately considered.

While Sect.~\ref{sec:gravlens} reviews the basic properties of gravitational lensing,
Sect.~\ref{sec:adaptive} presents the adaptive contouring algorithm and Sect.~\ref{sec:area}
shows how to determine the magnification of an extended source star from the image contour line.
Sect.~\ref{sec:binary} discusses the critical curves and caustics of a binary point-mass lens in
order to derive an algorithm for finding a point inside an image stretching over a critical curve.
Example light curves are shown in Sect.~\ref{sec:lightcurves} before the paper concludes with
a short summary in Sect.~\ref{sec:summary}.

\section{Gravitational Lensing}
\label{sec:gravlens}

Light received from a source object at distance $D_\rmn{S}$ is bent due to the gravitational field of a thin sheet of matter at a distance $D_\rmn{L}$ with surface mass density $\Sigma({\vec \xi}')$ by the angle
\citep{Schneider:theory}
\begin{equation}
\hat {\vec \alpha}(\vec \xi) = \frac{4G}{c^2} \int
\Sigma({\vec \xi}')\,\frac{\vec \xi -{\vec \xi}'}
{|\vec \xi -{\vec \xi}'|^2}\,\rmn{d}^2{\vec \xi}'\,.
\end{equation}
For $\vec \beta$ denoting the true source position angle and $\vec \theta$ the apparent position angle of its observed images, this implies the 
{\em lens equation}
\begin{equation}
\vec y (\vec x) = \vec x - \vec \alpha(\vec x)\,,
\label{eq:lenseq}
\end{equation}
where $\vec x = \vec \theta/\theta_0$, $\vec y = (D_\rmn{L}/D_\rmn{S})\,(\vec \beta/\theta_0)$,
and
\begin{equation}
\vec \alpha(\vec x) = \frac{4G}{c^2}\,\frac{D_\rmn{S}-D_\rmn{L}}{D_\rmn{L} D_\rmn{S}}\,\frac{1}{\theta_0^2}\,\int \kappa({\vec x}')\,
\frac{\vec x -{\vec x}'}
{|\vec x -{\vec x}'|^2}\,\rmn{d}^2{\vec x}'
\end{equation}
with $\kappa({\vec x}') = (D_\rmn{L}\theta_0)^2\,\Sigma(D_\rmn{L} \theta_0
{\vec x}')$. It provides a surjective mapping of the image position $\vec x$
to the source position $\vec y$, but the lack of injectivity means that a source 
may have more than just one image. 

At {\em critical points} ${\vec x}_\rmn{c}$, defined by
\begin{equation}
\det \left(\frac{\partial \vec y}{\partial \vec x}\right)\left({\vec x}_\rmn{c}\right) = 0\,,
\label{eq:defcrit}
\end{equation}
two images merge, so that the number of images changes if and only if the source passes a {\em caustic
point} ${\vec y}_\rmn{c} = {\vec y}({\vec x}_\rmn{c})$. In general, critical points and caustic points
form closed curves, known as {\em critical curves} and {\em caustics} unless these degenerate into
a point in special cases.
The conservation of surface brightness by gravitational lensing, i.e.\
$I(\vec x) = I[\vec y(\vec x)]$, implies that
the total magnification of a point source is given by
\begin{equation}
A(\vec y) = \sum_{\{{\vec x}^{(i)}|\vec y = \vec y({\vec x}^{(i)})\}}
\left[\det \left(\frac{\partial \vec y}{\partial \vec x}\right)\left({\vec x}^{(i)}\right)\right]^{-1}
\end{equation}
so that it diverges if the source comes to lie on a caustic.

For an extended source, as a consequence of Liouville's theorem, 
source and lens contours of same brightness correspond.
This means that 
if a source contour is described by an implicit function
$F(\vec y; \vec p) = 0$, with $\vec p$ being a parameter vector specifying
the contour, then all its image contours are given by
$F[\vec y(\vec x); \vec p] = 0$. 
Therefore, a contour plot provides image contour lines without inversion of the lens equation \citep{SK87}.

\section{Adaptive Contouring}
\label{sec:adaptive}

\subsection{Data structure and algorithm}

A search grid in the image plane needs to be large enough to cover all images and dense enough, so that
no images or holes in the images are missed. While fulfilling these two
condition, the grid resolution should be kept as low as possible.
Obviously, an adaptive grid that just provides higher resolution in regions
where this is required turns out to be superior to a fixed-resolution grid.

Such an adaptive grid can be built by hierachically nesting squares that
represent parts of the image plane. In order to approximate the position 
of the contour line, a ray is shot to the corresponding image position
given by the defined mapping, Eq.~(\ref{eq:lenseq}), 
and it is noted whether the corresponding true
source position falls inside or outside the source contour.  
In this respect, the proposed algorithm resembles the 
ray-shooting approach \citep{KRS:rayshooting,SchneiWei:AGN}, 
but the indicator of whether the light ray hits the source is kept with
the image position. In fact, with 
the conservation of surface brightness, as pointed out in
Sect.~\ref{sec:gravlens}, 
the image position has the same relation (inside or outside) with respect to the image contour as the corresponding source position to the source contour.

\begin{figure}
\includegraphics[width=84mm]{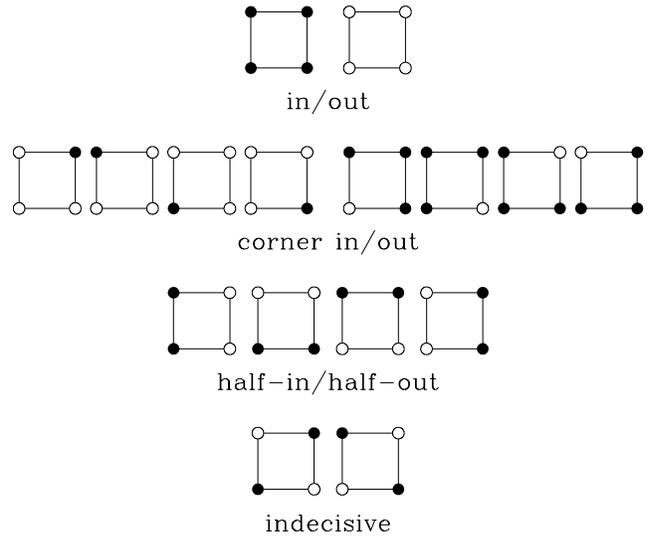}
\caption{The 16 different elementary squares, defined by whether each
of the corners is inside or outside the contour line, where 'inside' is marked
by a filled black circle and 'outside' is marked by an open circle (filled white).
While in- or out-squares are considered fully inside or outside the contour line and
therefore of no further interest, the contour line passes through squares
of type 'corner in/out' or 'half-in/half-out'. All four edges of squares
with corners alternating between inside and outside along their boundary 
are cut by the contour line, while it is not clear which edges the contour
line runs between, making such squares indecisive.}
\label{fig:square_type}
\end{figure}

A square with the inside/outside indicators for its corners constitutes the elementary data structure,
out of which all relevant information about the image plane is constructed. Being characterized
by its four indicators, there are $2^4 = 16$ types of elementary squares,
shown in Fig.~\ref{fig:square_type}.
12 of these elementary squares, namely those of type 'corner in/out' or 'half-in/half-out',
define part of the contour
line, which crosses two edges of the square whose corners have
different status. If two opposite
corners are found to be inside and the other two corners outside, the contour needs to cross all four edges, but it is not clear which edges
it connects. Therefore, such indecisive squares need to be subdivided 
in order to improve the resolution (this is rule 1 below).

The length of an 
edge of a square of depth $k$ is $2^{-k} a$, where $a$ denotes
a unit size. Squares are nested so that
each square $S^k_{mn}$ of depth $k$ either contains 4 subsquares of
depth $k+1$ each
sharing a different corner with $S^k_{mn}$, or $S^k_{mn}$ does not have
any subsquare.
The minimal and initial data structure consists of a square of depth $-1$
with 4 subsquares of depth 0.

\begin{figure}
\includegraphics[width=84mm]{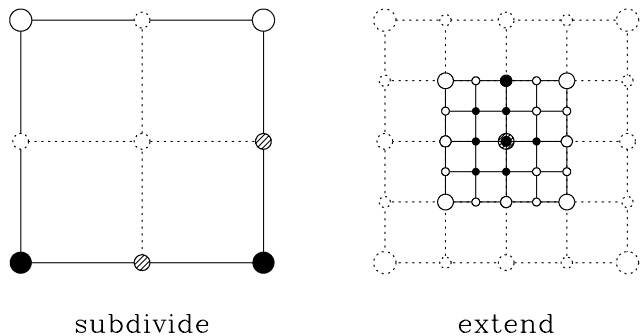}
\caption{The two operations {\tt subdivide} and {\tt extend}. The 
elements that are being added are shown with dotted lines or shaded
areas. The size of the inside/outside indicators reflects the depth of the
respective square.}
\label{fig:operations}
\end{figure}

The version of the adaptive contouring algorithm presented here uses 
just two operations on the data structure, {\tt subdivide} and {\tt extend}, illustrated in Fig.~\ref{fig:operations}. While {\tt subdivide}
creates 4 sub-squares of depth $k+1$ to a specified square of depth $k$, {\tt extend} enlarges
the range covered by the adaptive search grid by creating a new top-level grid with the same centre as
the previous one but twice the length of an edge. 
The following conditions require invocation of the operations
{\tt extend} (condition $\cal E$) or 
{\tt subdivide} (conditions ${\cal S}_1$ to ${\cal S}_3$):
\begin{itemize}
\item{{\em Condition} $\cal E$:} 
A point at the boundary of the outermost square is found to be inside 
\item {{\em Condition} ${\cal S}_1$:}
Square is indecisive, i.e.\ it has alternating inside/outside corners
along its boundary
\item {{\em Condition} ${\cal S}_2$:}
Adjoining corners of a square have same status, but status changes
among evaluated points (corners of subsquares) along the connecting edge 
\item {{\em Condition} ${\cal S}_3$:}
Square of depth smaller than a specified $k_\rmn{min}$ is of type 'corner in/out' or 'half-in/half-out' 
\end{itemize}

The operations following conditions $\cal E$, ${\cal S}_1$ and
${\cal S}_2$ eliminate all problematic cases
and need to be executed for squares of all depths before the contour line
can be determined, whereas calling {\tt subdivide} on all squares
fulfilling condition ${\cal S}_3$ leads to all squares that contain
the contour line having a depth of at least $k_\rmn{min}$.
Application of {\tt extend} can cause condition ${\cal E}$
only to arise in the new surrounding square. Similarly, the application
of {\tt subdivide} can cause condition ${\cal S}_1$ only to arise
in subsquares. In contrast, adjoining squares need to be 
checked for condition ${\cal S}_2$, while any of the conditions
may arise in new subsquares created on the application
of {\tt extend}. For this reason, {\tt extend} or {\tt subdivide}
are executed recursively if conditions ${\cal E}$ or ${\cal S}_1$ arise,
whereas if modifications lead to squares    
fulfilling conditions ${\cal S}_2$ or ${\cal S}_3$,
these are put on stacks specific to the condition and the depth of the
respective square. Since squares for any depth need to be checked
for condition ${\cal S}_2$, a square fulfilling it needs to be placed
on the stack even if it has alreadly been placed on a stack related 
to condition ${\cal S}_3$. This can leave squares on stacks for
condition ${\cal S}_3$ that have already been subdivided. If such squares are encountered,
they must be ignored and removed from the stack.

\begin{figure}
\includegraphics[width=84mm]{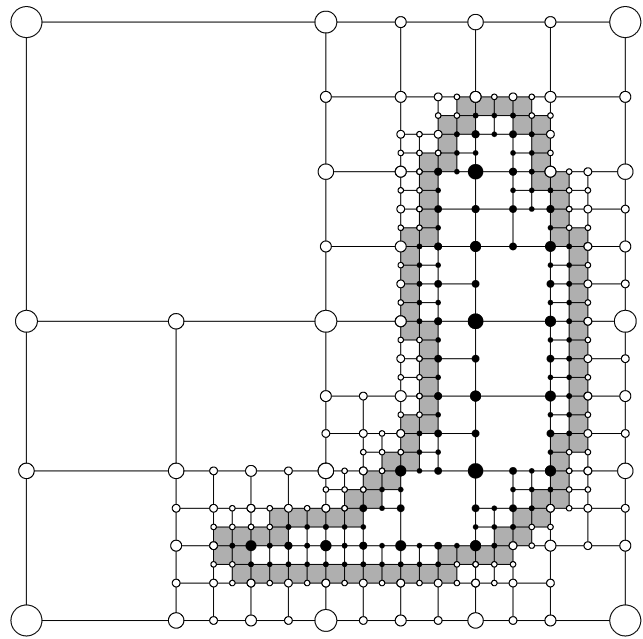}
\caption{Representation of contour at a given resolution depth
by set of nested squares after 
the rules of the adaptive contouring algorithm have been applied.
Filled black circles mark corners that are inside the contour, while open (filled white) circles
mark those outside the contour. Their size reflects the depth of the outermost square this corner
belongs to. The squares of types 'corner in/out' or 'half-in/half-out' (see Fig.~\ref{fig:square_type}) contain the
contour line and are shaded in grey.}
\label{fig:scan0}
\end{figure}

In order to find the contour line for a depth $k_\rmn{min}$,
from the top level to the depth $k_\rmn{min}-1$, squares are taken from the
corresponding stack related to condition ${\cal S}_3$ and the operation ${\tt subdivide}$ is executed.
This might cause conditions ${\cal E}$ and ${\cal S}_1$ to arise and 
corresponding {\tt extend} or {\tt subdivide} operations being carried out, 
while squares fulfilling conditions ${\cal S}_2$ are placed on a 
respective stack. 
Subsequently, all stacks with squares fulfilling ${\cal S}_3$ are being cleared by pulling the respective squares
from them and executing the {\tt subdivide} operation. If the application
of {\tt extend} for squares
fulfilling condition ${\cal E}$ or of {\tt subdivide} for squares
fulfilling condition ${\cal S}_2$ or were carried out, this might 
have created squares fulfilling condition ${\cal S}_3$, so that the full 
procedure needs to be repeated until this is not the case.
After all these operations have been applied, the stacks with squares
fulfilling condition ${\cal S}_3$ with depth $k \geq k_\rmn{min}$ hold
the squares from which to determine the contour line.
Fig.~\ref{fig:scan0} illustrates the representation of the image plane by the nested squares that results from applying the adaptive contouring
algorithm for a given resolution depth $k_\rmn{min}$.

As pointed out in detail in Sect.~\ref{sec:area},
The desired magnification of the source star can be obtained 
from an evaluation of approximated contour lines for a series of
increasing $k_\rmn{min}$ until a desired accuracy is reached.
Rather than subdiving all squares
that constitute the contour line to a certain given depth, it would be advantageous
to pick those squares that
are expected to provide the largest gain in improving the estimate of the
source magnification. However, this would require some advanced bookkeeping.

\subsection{Finding all images and holes}
The essential need to find all segments of the contour line can be
met by inserting a representation of at least one point inside each
image and each hole. A point that falls exactly onto a corner is represented directly
in the data structure. Otherwise, one can use the 
fact that with any point that is not exactly on the contour line,
there is always a region around it that is either inside or outside.
A point can then be represented by the surrounding region
which can take the form of a single in- or out-square if the point is inside a square and by
two adjoining in- or out-squares if the point falls on their common edge.
If 1) for each region enclosed by a contour line, at least one point is represented
by an inside corner or one or two in-squares, 
2) at least one point inside each hole (i.e.\ a region surrounded by a contour
line that is not enclosed) is represented by an outside corner or one or two out-squares
and 3) the representing squares do not overlap, the adaptive contouring
algorithm is guaranteed to find all segments of the contour line,
so that no enclosed region or hole is missed.

The condition for the representing squares not to overlap can be met by
determining a maximal size for the square surrounding each point to be inserted.
A simple choice is to consider touching squares of the same size
around this point and each of the other points, use the minimal size and round it 
to the next smaller $2^{-k_i} a$. 
If the point to be inserted into the data structure does not fall into the 
top-level square, one needs to call {\tt extend} until this
requirement is met. Subsequently, starting at the top-level square,
while the depth is smaller than the minimal depth $k_i$ 
or not all corners have the right status
(inside for images, outside for holes), one needs to branch
into the larger-depth subsquare that contains the point.
If no subsquares exist, these must be created by a {\tt subdivide} operation.
If the point to be inserted falls onto an existing corner, one can stop,
whereas if the point falls onto an edge,
the search needs to be split into searches inside a square right and left
(for a horizontal edge) or above and below (for a vertical edge). Such a
split becomes necessary at most once.

\subsection{Image contours for binary point-mass lenses}
\label{sec:binadap}

Let us consider the special case of a lens of total mass $M$, composed of two point lenses with
mass fractions $m_1$ and $m_2 = 1-m_1$ that are separated by the angle
$d\,\theta_\rmn{E}$, where
\begin{equation}
\theta_\rmn{E} = \sqrt{\frac{4GM}{c^2}\,
\frac{D_\rmn{S}-D_\rmn{L}}{D_\rmn{L} D_\rmn{S}}}\,.
\label{eq:Einsteinang}
\end{equation}
With the choice $\theta_0 = \theta_\rmn{E}$, the lens equation,
Eq.~(\ref{eq:lenseq}), then 
reads
\begin{eqnarray}
\vec y (\vec x) &  = & \vec x - m_1 \frac{\vec x + (1-m_1) (d,0)}{|\vec x + (1-m_1) (d,0)|^2} \;- \nonumber \\
& & \quad -\;
(1-m_1) \frac{\vec x - m_1(d,0)}{|\vec x - m_1 (d,0)|^2}\,,
\label{eq:binlenseqcom}
\end{eqnarray}
where the origin of the coordinate system coincides with the centre of mass
of the binary-lens system and the line connecting the two components is along the $y_1$-axis. Such a binary point-mass lens is not only a fair
model for a stellar binary but also for an isolated star orbited by planets, where usually the effect of just a single planet dominates.

If an extended source is completely outside the caustics, its image 
contours surround three images, whereas these surround five images if
the extended source is completely inside. Every point inside the source
has a point image inside one of the images of the extended source, so that
the images of any point inside the source, in particular the source
centre,  provide a seed for finding the
image contour lines as pointed out in the previous subsection.
These can be found straightforwardly by solving a fifth-order complex polynomial \citep{WM95:fifth}.
If part of the source is inside and part of it outside, the number of images differs among its enclosed points, so that between one and four images of the extended source can occur, where some contour lines cross 
critical curves, so that the enclosed images of the source contain
more than a single image of points within the source that are inside
the caustic. Finding a point inside the caustic and determining its images
would provide all five necessary seeds. While this is not trivial, one
can take a slightly different approach. As long as the source encloses
a cusp, the contour line encloses at least one image of any point inside
the caustic, so that the at least three images of the source centre
are sufficient as seeds. The only remaining case is that the source
extends over a single fold caustic, so that one of its images stretches 
over a critical curve with no position inside mapping to the source centre.
A point inside such an image can be found as a point on the critical curve that maps to a local minimum of the distance between source centre and caustic. 
Finding such a point inside requires some sophisticated analysis of the
critical curves and caustics of the underlying binary lens.
This is pointed out in detail in Sect.~\ref{sec:binary}.
Seeds for holes are found easily for a binary point-mass lens since these
need to include one of the positions of the two point-mass constituents.
With the lens equation, Eq.~(\ref{eq:binlenseqcom}), having poles there, 
it is sensible to ensure that these positions are defined to be outside 
any closed contour.

\begin{figure}
\includegraphics[width=84mm]{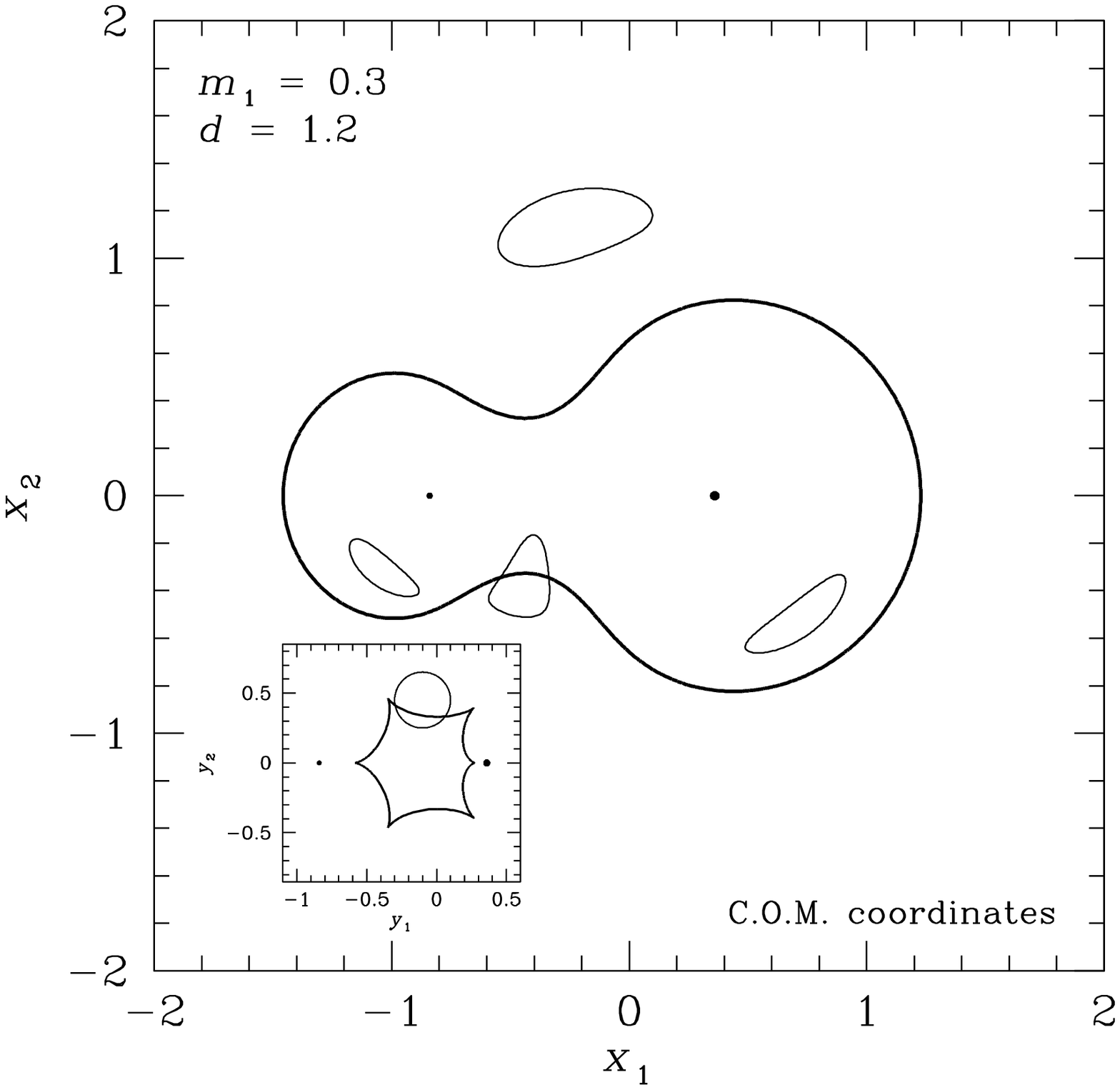}\\
\includegraphics[width=84mm]{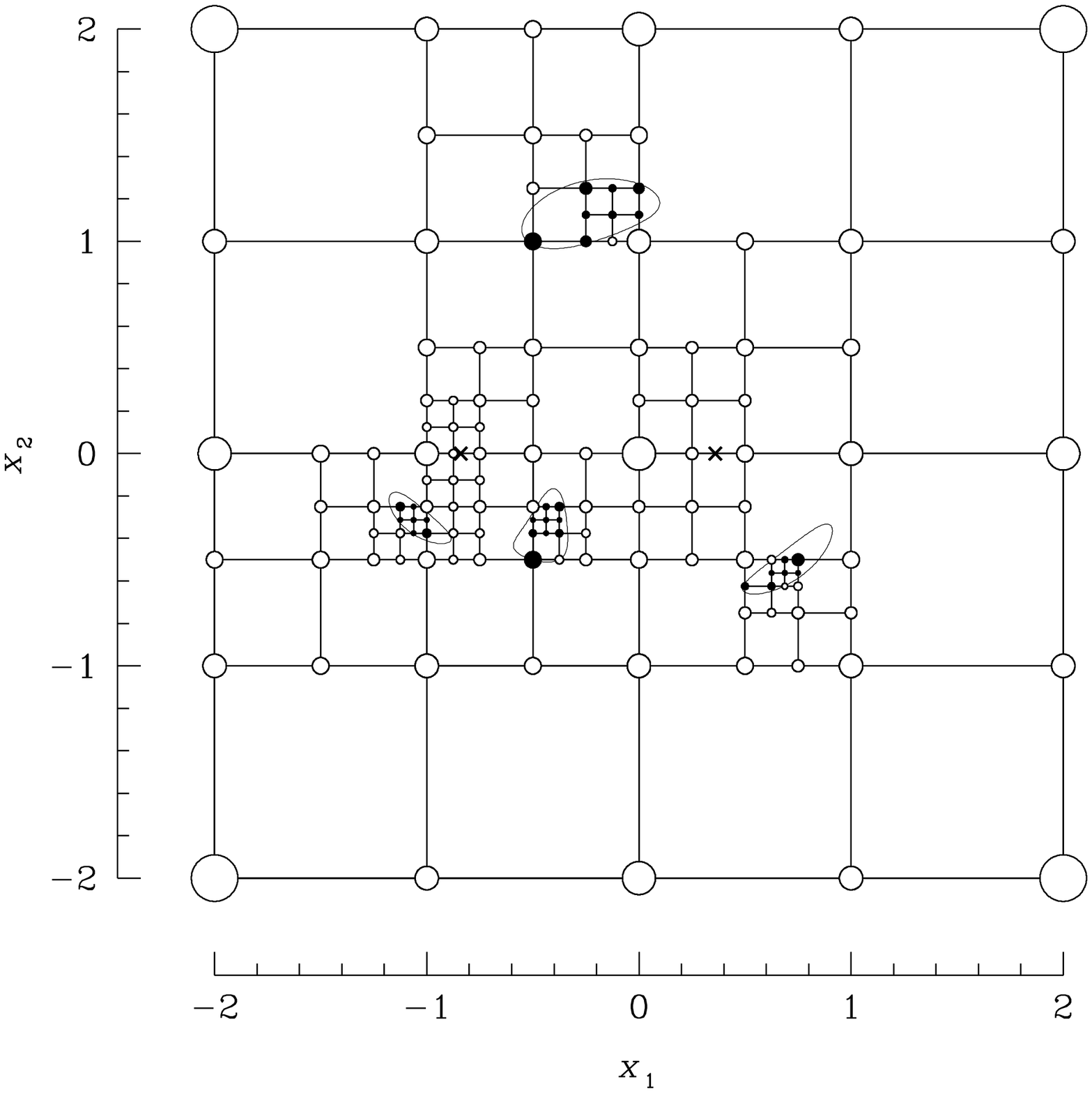}
\caption{(top) Image contours for a circular source star resulting from the deflection of light by the gravitational field of a binary lens for an example configuration. The coordinates denote angles on the sky in multiples of the angular Einstein radius $\theta_\rmn{E}$, defined by Eq.~(\ref{eq:Einsteinang}), where the coordinate origin coincides with the centre of mass of
the binary-lens system. The separation of the two point masses with fractional mass $m_1 = 0.3$ (left) and $1-m_1 = 0.7$ (right) in units of the angular Einstein radius $\theta_\rmn{E}$ has been chosen as $d = 1.2$.
The thick line shows the critical curve. The inset shows the caustic along with the source contour, where the source
centre is located at $\vec y = (-0.1,0.45)$ and its angular radius in units of $\theta_\rmn{E}$ is $\rho_\star = 0.2$. 
(bottom) Set of nested squares with inside/outside indicators constituting the data structure that represents the 
image plane after seeds for the image and the hole positions have been inserted. Image seeds were the three images 
of the source centre and a point on the critical curve that maps to
a local minimum of the distance between source centre and caustic.
Hole seeds were the positions of the two point masses, namely
$(-0.84,0)$ and $(0.36,0)$, marked as crosses. In lightface, the image contour lines are indicated.
The size of the inside/outside indicators (black for inside, white for outside) reflect the depth of the respective 
outermost square.}
\label{fig:scan}
\end{figure}

Fig.~\ref{fig:scan} shows the data structure after the seed points
have been inserted for an example configuration, for which
$m_1 = 0.3$, $d = 1.2$ and the centre of a circular source 
of angular radius $\rho_{\star}\,\theta_\rmn{E}$,
where $\rho_{\star} = 0.2$, is located at $\vec y = (-0.1,0.45)$ along
with the image contour lines, the critical curve, and the source 
contour with the caustic.
Around the positions of the two point masses, namely $(-0.84,0)$ and $(0.36,0)$, marked by crosses, two out-squares representing holes have
been inserted (these positions falling onto an edge), whereas the three images of the source centre have been surrounded by in-squares. However, there is a fourth
image stretching over a critical curve, where a seed has been given by a local minimum of the distance between the source
centre and the caustic, as described in detail in Sect.~\ref{sec:getcritimg}.

\section{Image area and magnification}
\label{sec:area}
The conservation of surface brightness by gravitational lensing
implies that for uniformly
bright sources, the magnification equals the ratio between the total area
of the images and that of the unaffected source. This means that the
problem of calculating the magnification of a uniformly bright source reduces
to that of determining the area of its images.

As exploited by \citet{Do95:Num,Do98:NumSrc}, 
Green's theorem allows to transform an integral over a region $R$ into an integral
over its boundary $\partial R$. Namely, for two functions $P(x_1,x_2)$ and $Q(x_1,x_2)$ that are
continous in $R$ and whose partial derivatives $\partial P/\partial x_2$ and
$\partial Q/\partial x_1$ are also continuous in $R$,
\begin{equation}
\int\limits_{R} \left(\frac{\partial Q}{\partial x_1} - \frac{\partial{P}}{\partial x_2}\right)\,
\mathrm{d}x_1\,\mathrm{d}x_2 =
\int\limits_{\partial R} P\,\mathrm{d}x_1 + Q\,\mathrm{d}x_2\,.
\end{equation}
In particular, for the area of $R$, one has ${\partial Q}/{\partial x_1} - {\partial{P}}/{\partial x_2} = 1$,
which is fulfilled with the choice $P(x_1,x_2) = -\frac{1}{2} x_2$ and $Q(x_1,x_2) = \frac{1}{2} x_1$, so that
the magnification of a uniformly bright circular source star with angular radius $\rho_\star\,\theta_\mathrm{E}$ reads
\begin{equation}
A({\vec y}^{(0)},\rho_\star) = \frac{1}{2\upi\,\rho_\star^2}\,\int\limits_{\partial R} x_1\,\rmn{d}x_2 - x_2\,\rmn{d}x_1\,,
\label{eq:maguniform}
\end{equation}
where $\partial R$ denotes the union of the contours of all its images.

An approximation of the contour line can be obtained from an inspection of all squares
of types 'corner in/out' or 'half-in/half-out'.
The intermediate value theorem implies that the continuous function
$F[\vec y(\vec x),\vec p]$ restricted to the edge of a square has at least one root
if the signs at the corners
differ, which means that the contour line has to cross the edge. 
For elementary square of types 'corner in/out' or 'half-in/half-out', see
Fig.~\ref{fig:square_type}, there are two such edges whose crossings with
the image contour line define its entry to and exit from the respective
square. Usually contour plot algorithms determine the cut of the contour line
with the edge by linear interpolation using the values of $F[\vec y(\vec x),\vec p]$ at the corners.
However, since the images are dominated by convex 
contours, this choice systematically leads to an underestimation of the 
enclosed area. From practice, it has emerged that simply using the 
midpoint between the corners provides an estimate that on average deviates
less from the true image area for moderate precisions.
It might happen that the image contour line runs through squares of different
depth. For adjacent squares of different depth joining at an edge,
the contour line has to be put through the edge of the square with larger 
depth, i.e.\ the smaller one.  

Let us consider a circular source with angular radius $\rho_\star\,\theta_\rmn{E}$ whose
center is at ${\vec y}^{(0)}$, so that $\vec p = ({\vec y}^{(0)}, \rho_\star)$ constitutes
the respective parameter vector.
For the square denoted by the running index $i$, let 
$(x_1^{(i)},x_2^{(i)})$ and $(x_1^{(i+1)},x_2^{(i+1)})$ be the adopted intersections of the edges
with the contour line. A symmetric discretization of Eq.~(\ref{eq:maguniform}) then yields
\begin{eqnarray}
A({\vec y}^{(0)}, \rho_\star) & = & \frac{1}{4 \pi \rho_\star^2}\,\sum_{i=1}^{n} 
\left[(x_1^{(i)}+x_1^{(i+1)})\,(x_2^{(i+1)}-x_2^{(i)}) \,-\right. \nonumber \\
& & \left.-\,(x_2^{(i)}+x_2^{(i+1)})\,(x_1^{(i+1)} - x_1^{(i)})\right] \nonumber \\
& = & \frac{1}{4 \pi \rho_\star^2}\,\sum_{i=1}^{n} \left( x_1^{(i)}  x_2^{(i+1)}
-  x_2^{(i)}  x_1^{(i+1)}\right)\,.
\label{eq:magnubdiscrete}
\end{eqnarray}

The absolute error in the area cannot exceed the sum of the area of all squares through which the contour line runs,
but it can be expected to be much smaller than that. While the ratio $\varepsilon$ between these squares and the approximated enclosed area gives some measure for the quality of the approximation, it vastly overestimates the typical error. However, a suitable value for $\varepsilon$ can be chosen from testing how its variation affects the final result.

For a general surface brightness $I[{\vec y}(\vec x); {\vec y}^{(0)}, \rho_\star]$, one finds
\begin{eqnarray}
& & \hspace*{-2em} 
A({\vec y}^{(0)}, \rho_\star) \;= \;\frac{1}{4 \pi \rho_\star^2} \sum_{i=1}^{n}
\Bigg\{\left[J_1\left(x_1^{(i)},\frac{x_2^{(i)}+x_2^{(i+1)}}{2}\right)\right. \; + \nonumber \\
& & \hspace*{-2em} \qquad \quad +\;\left.J_1\left(x_1^{(i+1)},\frac{x_2^{(i)}+x_2^{(i+1)}}{2}\right)\right] \left[x_2^{(i+1)}-x_2^{(i)}\right] \;- \nonumber \\
 & & \hspace*{-2em} \quad -\; \left[J_2\left(\frac{x_1^{(i)}+x_1^{(i+1)}}{2},x_2^{(i)}\right)\right.\;+ \nonumber \\
& & \hspace*{-2em} \qquad \quad +\;\left.J_2\left(\frac{x_1^{(i)}+x_1^{(i+1)}}{2},x_2^{(i+1)}\right)\right] \left[x_1^{(i+1)}-x_1^{(i)}\right]\Bigg\} 
\end{eqnarray}
where
\begin{eqnarray}
J_1(x_1,x_2) & = & \overline{I}^{-1} \int\limits_0^{x_1} I[{\vec y}(x'_1,x_2); {\vec y}^{(0)}, \rho_\star]\,\rmn{d}x'_1\,, \nonumber\\
J_2(x_1,x_2) & = & \overline{I}^{-1} \int\limits_0^{x_2} I[{\vec y}(x_1,x'_2); {\vec y}^{(0)}, \rho_\star]\,\rmn{d}x'_2\,.
\end{eqnarray}
and the brightness of the source, centered at ${\vec y}^{(0)}$ reads
\begin{equation}
I[{\vec y}(\vec x); {\vec y}^{(0)}, \rho_\star] = \overline{I} \times \left\{
\begin{array}{l}
\xi\left[\frac{|{\vec y}(\vec x)-{\vec y}^{(0)}|}{\rho_\star}\right] \\
\qquad  \mbox{for} \quad
|{\vec y}(\vec x)-{\vec y}^{(0)}| \leq \rho_\star \\
0 \\
\qquad 
\mbox{for} \quad
|{\vec y}(\vec x)-{\vec y}^{(0)}| > \rho_\star
\end{array}
\right.
\end{equation}
with $\overline{I}$ being the average surface brightness and $\xi(\rho)$ being the characteristic
profile function depending on the fractional radius $\rho$.

However, $I[{\vec y}(\vec x); {\vec y}^{(0)}, \rho_\star]$
jumps at the source boundary $|{\vec y}(\vec x)-{\vec y}^{(0)}| = \rho_\star$ if $\xi(1) \neq 0$, and its derivative
at that position jumps if $(\rmn{d}\xi/\rmn{d}\rho)(1) \neq 0$. 
Similarly to the definition made by \citet{Do98:NumSrc} for a
linear limb-darkening profile, this can be avoided for general
monotically falling $\xi(\rho)$ 
by defining
\begin{equation}
\tilde{I}[{\vec y}(\vec x); {\vec y}^{(0)}, \rho_\star] = 
\overline{I} \times \left\{
\begin{array}{l}
\xi\left[\frac{|{\vec y}(\vec x)-{\vec y}^{(0)}|}{\rho_\star}\right] 
+\xi(0)- 2\xi(1)\\
\qquad  \mbox{for} \quad
|{\vec y}(\vec x)-{\vec y}^{(0)}| \leq \rho_\star \\
\xi(0) - \xi\left[\frac{\rho_\star}{|{\vec y}(\vec x)-{\vec y}^{(0)}|}\right]\\
\qquad 
\mbox{for} \quad
|{\vec y}(\vec x)-{\vec y}^{(0)}| > \rho_\star
\end{array}
\right.\!\!,
\end{equation}
for which $\tilde{I}[{\vec y}(\vec x); {\vec y}^{(0)}, \rho_\star] \geq 0$. With the value of $I[{\vec y}(\vec x) {\vec y}^{(0)}, \rho_\star]$ for $|{\vec y}(\vec x)-{\vec y}^{(0)}| > \rho_\star$ not affecting the integral, the 
magnification of the source can then be determined as
\begin{eqnarray}
A({\vec y}^{(0)}, \rho_\star) & = & \frac{1}{\upi \rho_\star^2}
\int \frac{I[{\vec y}(\vec x); {\vec y}^{(0)}, \rho_\star]}{\overline{I}}\,\rmn{d}x_1\,\rmn{d}x_2 \nonumber \\
 & = & \frac{1}{\upi \rho_\star^2} \Big\{
\int \frac{\tilde{I}[{\vec y}(\vec x); {\vec y}^{(0)}, \rho_\star]}{\overline I}\,\rmn{d}x_1\,\rmn{d}x_2 \; - \nonumber\\
& & \quad - \;[\xi(0) - 2 \xi(1)] \int \rmn{d}x_1\,\rmn{d}x_2\Big\}\,.
\end{eqnarray}

\section{Critical curves of binary lenses}
\label{sec:binary}

\begin{figure}
\includegraphics[width=84mm]{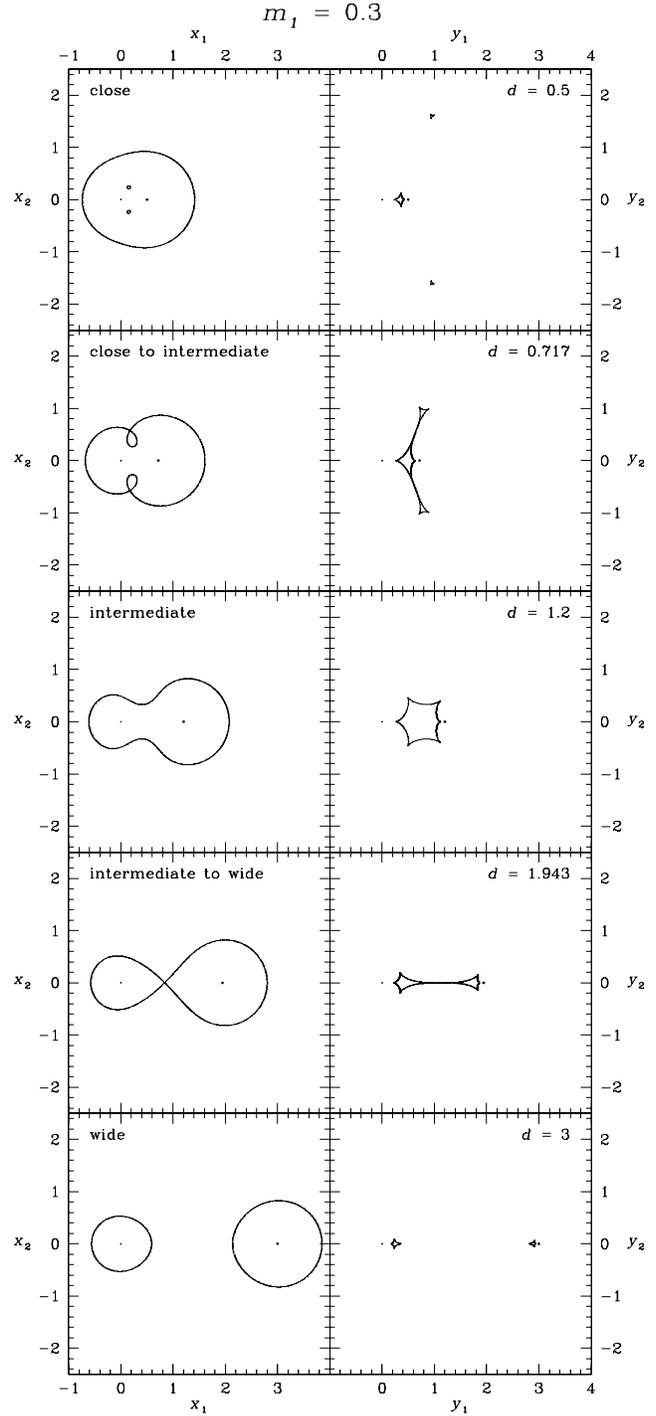}
\caption{The three different topologies for the critical curves (left)
and caustics (right) of binary point-mass lenses, illustrated for
$m_1 = 0.3$, along with the transitions, which
occur at separation parameters
$d_\rmn{c} = 0.717$ or $d_\rmn{w} = 1.943$. The choice
of the coordinate system corresponds to Eq.~(\ref{eq:binlenseq}).
Small dots indicate the positions of the two lens objects.}
\label{fig:cctopologies}
\end{figure}

\subsection{Topology}
For studying the critical curves of a binary lens, it is favourable to
center the coordinate system on the component with mass fraction $m_1$
rather than on the centre of mass as was done in Sect.~\ref{sec:binadap},
while keeping the line connecting the two components is along the $y_1$-axis. This means that for the two point lenses with
mass fractions $m_1$ and $m_2 = 1-m_1$ that are separated by the angle
$d\,\theta_\rmn{E}$ one finds the lens equation
\begin{equation}
\vec y (\vec x) = \vec x - m_1 \frac{\vec x}{|\vec x|^2} -
(1-m_1) \frac{\vec x - (d,0)}{|\vec x - (d,0)|^2}\,,
\label{eq:binlenseq}
\end{equation}
where the coordinate system has been centered on the primary and 
the line connecting the two components is along the $y_1$-axis.

For every mass fraction $m_1$, the topology of the critical curves
and the corresponding caustics undergo transitions at characteristic
separations $d_\rmn{c}$ and $d_\rmn{w}$,
given by \citep{Erdl}
\begin{eqnarray}
m_1 (1-m_1) & = & \frac{1}{d_\rmn{c}^8}\,\left(\frac{1-d_\rmn{c}^4}{3}\right)^3 \,, \nonumber \\
d_\rmn{w} &  = & \left[m_1^{1/3} + (1-m_1)^{1/3}\right]^{3/2}\,.
\end{eqnarray}
This results in the three cases
of close binaries ($d < d_\rmn{c}$), intermediate binaries
($d_\rmn{c} \leq d \leq d_\rmn{w}$), and wide binaries
($d > d_\rmn{w}$).
For an equal-mass binary in particular, the transitions between the
topologies occur at $d_\rmn{c} = 1/\sqrt{2}$
and $d_\rmn{w} = 2$ \citep{SchneiWei:twomass}. The three different topologies
and their transitions are illustrated in Fig.~\ref{fig:cctopologies}, where
the case $m_1 = 0.3$ has been used as example. For $d \to \infty$, the critical curves
tend towards circles with radii $\sqrt{m_1}$ or $\sqrt{1-m_1}$ around each of the individual
point mass, whereas the corresponding caustics are diamond-shaped with four cusps each, where
two of these are on the $y_1$-axis, onto which the components of the
lens binary fall, and tend towards points
at the position of each point mass. As the two point masses approach, the inner cusps on the $y_1$-axis
touch, as well as the corresponding critical points, so that the critical curve becomes a lying '8'. For intermediate binaries, i.e.\ $d_\rmn{c} < d <
d_\rmn{w}$, there is a single closed critical curve in the shape of the bone,
while the corresponding caustic contains 6 cusps, 2 of them being on
the binary axis. Towards smaller separations, the fold line bends in and 
touches itself for $d = d_\rmn{c}$, so that one outer and two inner closed
curves result for a close binary, where $d < d_\rmn{c}$. Correspondingly,
the two upper or lower fold branches next to the on-axis cusps touch and
the caustic detaches into a diamond-shaped 4-cusp caustic with two on-axis cusps and two
triangular 3-cusp caustic with all those cusps being off-axis.
As $d \to 0$, the outer critical curve tends to a circle of unit radius around
the centre of mass, the inner critical curves contract towards a point and
move towards the centre of mass. The outer critical curve
corresponds to the diamond-shaped caustic, which tends towards a point at the
centre of mass, whereas the triangular-shaped caustic, corresponding to the
inner critical curves, decrease more strongly in size and move towards 
infinity.

\subsection{Parametric representation}

Using polar coordinates $\vec x = r (\cos \varphi, \sin \varphi)$, 
\citet{Erdl} have pointed out that critical points have to fulfill
an equation of the form
\begin{equation}
a_2(r;d,m_1) \cos^2 \varphi + a_1(r;d,m_1) \cos \varphi + 
a_0(r;d,m_1) = 0\,,
\label{eq:solvephi}
\end{equation}
which can readily be solved for $\cos \varphi$ for any given $r$, so that
critical curves can be parametrized by the radial coordinate $r$ for
each branch defined by the different values of $\cos \varphi$. In order to
do so, however, one needs to find the ranges for $r$, for which 
critical points exist. The $\cos \varphi$-dependence gives two
values $\varphi^{(+)} \in (0,\upi)$ or $\varphi^{(-)} \in (\upi,2\upi)$,
where $\varphi^{(+)} + \varphi^{(-)} = 2 \upi$, for 
$-1 < \cos \varphi < 1$, which reflects the symmetry around the axis 
connecting the two lens components.

With 
\begin{eqnarray}
G & = & r^2 + d^2 \,, \nonumber \\
H & = & 2dr \,, \nonumber \\
g & = & 1 - \frac{m_1^2}{r^4} \,,\nonumber \\
h & = & \frac{2 m_1 (1-m_1)}{r^2}\,,
\end{eqnarray}
\cite{Erdl} found
\begin{eqnarray}
a_2(r;d,m_1) & = & H^2 g - 2 d^2 h\,, \nonumber \\
a_1(r;d,m_1) & = & H(h-2Gg)\,, \nonumber \\
a_0(r;d,m_1) & = & G (Gg-h)+2 d^2 h - (1-m_1)^2\,.
\end{eqnarray}
The existence of critical points requires the discriminant of the 
quadratic equation for $\cos \varphi$, Eq.~(\ref{eq:solvephi}), to be non-negative,
i.e. $D \equiv a_1^2 - 4 a_0 a_2 \geq 0$ as well as
$-1 \leq \cos \varphi \leq 1$. With Eq.~(\ref{eq:solvephi}), one
finds the extreme values for $\cos \varphi$
to be assumed for $P_{-} \equiv a_2-a_1+a_0 = 0$ or
$P_{+} \equiv a_2+a_1+a_0 = 0$, respectively. This means that the number of solutions
for $\cos \varphi$ can only change for $D = 0$, $P_{-} = 0$, or $P_{+} = 0$,
so that the respective values for $r$ at given $(d,m_1)$ define minimal or maximal 
values of branches $(r,\varphi)$ of the critical curve.
For $a_2 = 0$, i.e.\ $r = m_1^{1/4}$ irrespective of $d$,
$C \equiv \cos \varphi$ has the unique solution $C = - a_0/a_1$ 
(for $a_2 = 0$ and $d > 0$, one always finds $a_1 \neq 0$)
with Eq.~(\ref{eq:solvephi}), while one finds
\begin{equation}
C_{\pm}  =  -0.5 [a_1\pm(a_1^2 - 4 a_0 a_2)^{1/2}]/a_2\,, 
\label{eq:branchphi}
\end{equation}
where $C_{+}$ and $C_{-}$ coincide for $D \equiv a_1^2 - 4 a_0 a_2 = 0$.

\begin{figure}
\includegraphics[width=84mm]{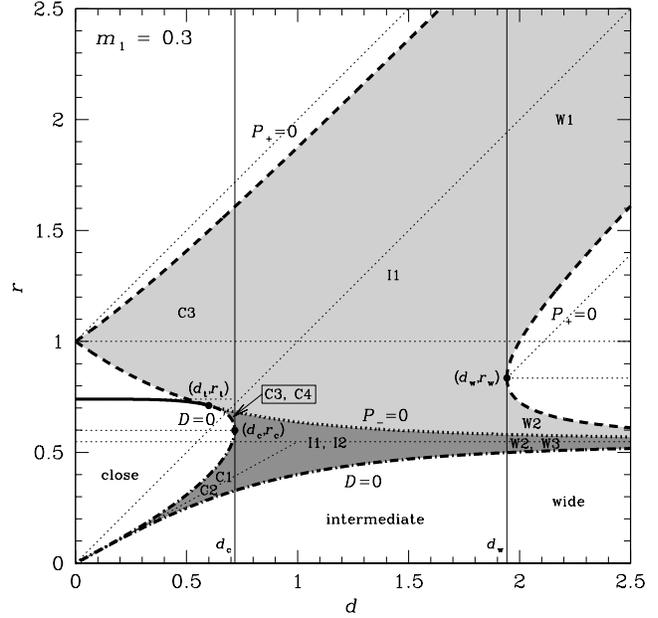}
\caption{Structure of the branches of the critical curve $(r,\varphi)$ indicated 
by the allowed ranges $[r_\rmn{min},r_\rmn{max}]$ for $r$ marked by thick lines, which
correspond to $D = 0$, $P_{-} = 0$, or $P_{+} = 0$. The line type reflects the
matching value of $\cos \varphi$, either $C_{+}$ or $C_{-}$, Eq.~(\ref{eq:branchphi}).
While dashed lines indicate a limit for $r$ with $C_+$, dotted lines
refer to a limit with $C_{-}$, and dash-dotted lines with both. The solid part of
the $D = 0$ contour refers to neither. Bracket intervals for the critical radii,
as listed in Table~\ref{tab:brackr}, are indicated by thin dotted lines.
The lighter shaded region marks the range for which a solution with $C_{+}$ exists,
whereas the darked shades region marks the range for which both solutions with either
$C_{+}$ or $C_{-}$ exist.
The classification of the branches as C1 to C4, I1 and I2, and W1 to W3 for
close, intermediate or wide binaries, is shown in Table~\ref{tab:branches}.
The transitions between the topologies are indicated by thin solid lines while
the points where the critical curves touch are marked as $(d_\rmn{c}, r_\rmn{c})$
or $(d_\rmn{w}, r_\rmn{w})$. The contours for $D=0$ and $P_{-} = 0$ touch
at $(d_\rmn{t}, r_\rmn{t})$.}
\label{fig:solstruct}
\end{figure}

Fig.~\ref{fig:solstruct} shows the allowed ranges for $r$ as a function of $d$
using $m_1 = 0.3$ as example. However, the structure of the solutions does not
depend on the adopted mass fraction. While for given
$(d,m_1)$, there is at least one and up to three solutions for $D = 0$, 
denoted in the following by $r_{D}^{(1)}$, $r_{D}^{(2)}$, $r_{D}^{(3)}$, and $P_{+} = 0$,
denoted by $r_{P_{+}}^{(1)}$, $r_{P_{+}}^{(2)}$, and $r_{P_{+}}^{(3)}$, there is
always just a single solution for $P_{-} = 0$, denoted by $r_{P_{-}}$.
The two additional solutions for $D = 0$ are only present
for close binaries, whereas the two additional solutions
for $P_{+} = 0$ exist for wide binaries.
These critical radii can be obtained by interval bisection using the initial
bracket intervals listed in Table~\ref{tab:brackr}, which are also indicated by
small dotted lines in Fig.~\ref{fig:solstruct}.
For $d = 0$, one finds $r_{P_{-}} = r_{P_{+}}^{(1)} = 1$ which reflects
the outer critical curve tendings towards the constant $r = 1$ (Einstein
circle), while $r_{D}^{(1)} = r_{D}^{(2)} = 0$ reflects the inner critical curves
contracting at the origin. Moreover, $r_{D}^{(3)} = m_1^{1/4}$ for $d=0$, but this is not 
a valid value, because $\cos \varphi$ is out of the allowed range.

\begin{table}
\caption{Bracket intervals for the critical radii defining the branches
of the critical curves of a binary point-mass lens.}
\label{tab:brackr}
\begin{tabular}{@{}lll}
\hline
Critical radius & Bracket interval & Separation \\
\hline
$r_{P_{-}}$ & $(m_1^{1/2},1]$ \\
$r_{P_{+}}^{(1)}$ & $(d,d+1)$ \\
$r_{P_{+}}^{(2)
}$ & $[d-d_\rmn{w}+r_\rmn{w},d)$ & $d > d_\rmn{w}$ \\
$r_{P_{+}}^{(3)}$ & $(m_1^{1/2},r_\rmn{w}]$ & $d > d_\rmn{w}$ \\
$r_{D}^{(1)}$ & $(0,\min\{m_1^{1/2}d,m_1^{1/2}\})$\\
$r_{D}^{(2)}$ & $(m_1^{1/2}d,r_\rmn{c}]$ & $d < d_\rmn{c}$ \\
$r_{D}^{(3)}$ & $[r_\rmn{c},m_1^{1/4}]$ & $d < d_\rmn{c}$ \\
\hline
\end{tabular}

\medskip
For given mass fraction $m_1$ and separation parameter $d$, critical
radii need to fulfill $P_{-} = 0$, $P_{+} = 0$, or $D = 0$. These
roots can be found by interval bisection of the respective bracket
interval.
$r_{D}^{(1)}$ and $r_{D}^{(2)}$ are separated by the asymptotic solution
$m_1^{1/2}d$ for $d \to 0$.
\end{table}

\begin{table}
\caption{Branches of the critical curves defined by the range of the radial coordinate
$r$ and the branch of $\cos \varphi$, defined by Eq.~(\ref{eq:branchphi}).}
\label{tab:branches}
\begin{tabular}{@{}lccl}
\hline
Branch & Range of $r$ & $\cos \varphi$ & Separation \\
\hline
C1 & $[r_{D}^{(1)}, r_{D}^{(2)}]$ & $C_{+}$ & $d < d_\rmn{c}$ \\
C2 & $[r_{D}^{(1)}, r_{D}^{(2)}]$ & $C_{-}$ & $d < d_\rmn{c}$ \\
C3 & $[r_{P_{-}}, r_{P_{+}}^{(1)}]$ & $C_{+}$ & $d \leq d_\rmn{t} < d_\rmn{c}$ \\
C3 & $[r_{D}^{(3)}, r_{P_{+}}^{(1)}]$ & $C_{+}$ & $d_\rmn{t} < d < d_\rmn{c}$ \\
C4 & $[r_{D}^{(3)}, r_{P_{-}}]$ & $C_{-}$ & $d_\rmn{t} < d < d_\rmn{c}$ \\
I1 & $[r_{D}^{(1)}, r_{P_{+}}^{1}]$ & $C_{+}$ & $d_\rmn{c} \leq d \leq d_\rmn{w}$ \\
I2 & $[r_{D}^{(1)}, r_{P_{-}}]$ & $C_{-}$ & $d_\rmn{c} \leq d \leq d_\rmn{w}$ \\
W1 & $[[r_{P_{+}}^{(2)},r_{P_{+}}^{(1)}]$ & $C_{+}$ & $d > d_\rmn{w}$ \\
W2 & $[r_{D}^{(1)},r_{P_{+}}^{(3)}]$ & $C_{+}$ &  $d > d_\rmn{w}$  \\
W3 & $[r_{D}^{(1)},r_{P_{-}}]$ & $C_{-}$ & $d > d_\rmn{w}$\\
\hline

\end{tabular}

\medskip
Branches C1--C4 refer to the close-binary case, I1/I2 to the intermediate
binary and W1--W3 to wide binaries, where transitions occur at
$d_\rmn{c}$ and $d_\rmn{w}$, which depend on $m_1$. At $d_\rmn{t}$,
the curves $D = 0$ and $P_{-} = 0$ touch, so that branch C4 only 
exists for $d_\rmn{t} < d < d_\rmn{c}$.
\end{table}

Table~\ref{tab:branches} lists the branches of the critical curve defined by
the allowed range $[r_\rmn{min},r_\rmn{max}]$ for $r$ and the value for
$\cos \varphi$, either $C_{+}$ or $C_{-}$ as defined by Eq.~(\ref{eq:branchphi}). 
While C1 and C2 correspond to the small two inner critical curves, 
C3 and C4 represent the outer critical curve. Branch C4 only exists 
for $d_\rmn{t} < d < d_\rmn{c}$, where $d_\rmn{t}$ is the separation for
which the curves $D = 0$ and $P_{-} = 0$ touch. These conditions
imply $a_2 = a_0$, so that $a_1 = 2 a_0$ or $a_1 = 2 a_2$.
For $d \to 0$, C3 tends to the Einstein circle, whereas C1 and C2 degenerate into points.
For $d = d_\rmn{c}$, the inner critical curves touch the outer one at the radius $r_\rmn{c} = (m_1 d_\rmn{c})^{1/3}$,
where C1 and C3 merge into I1, and C2 and C4 merge into I2. For the intermediate
binary, I2 is the branch near the object with mass fraction $m_1$ at
the coordinate origin, whereas I1 is the far branch.
For wide binaries, W1 gives the critical curve around the far object,
while W2/W3 surround the near object, which both tend to circles around
the respective object. At $d = d_\rmn{w}$,
the critical curves described by branches W1 and W2 merge at radius $r_\rmn{w} = (m_1 d_\rmn{w})^{1/3}$
into I2, whereas W3 and I2 are identical.

\subsection{Cusps and fold lines}
In order for the determinant of the Jacobian $(\partial \vec y/\partial \vec x)$ of the lens mapping to vanish, at least one of its eigenvalues must be zero. In two dimensions, singularities for which there are no non-zero eigenvalues
cannot be a generic feature. In fact, such singularities (known as umbilics) do not exist for any binary
point-mass lens \citep{SchneiWei:twomass,Erdl}. Therefore, we can restrict ourselves to the case of exactly one eigenvalue being zero for dicussing the critical curves of binary lenses. 

Since the critical curve is defined as the set of points where the Jacobian determinant
of the lens mapping vanishes, Eq.~(\ref{eq:defcrit}), the unit normal vector is
given as $\vec n = \vec D/|\vec D|$, where
$\vec D = \nabla \det \left(\frac{\partial \vec y}{\partial \vec x}\right)$. A corresponding
tangent vector then follows as ${\vec t}_{\vec x} = (-n_2,n_1)$, and 
${\vec t}_{\vec y} = \left(\frac{\partial \vec y}{\partial \vec x}\right) {\vec t}_{\vec x}$ is
a tangent to the caustic. Therefore, the caustic is a smooth curve with defined tangent direction
(fold singularity) as long as ${\vec t}_{\vec x}$ is not an eigenvector ${\vec e}_0$ of the Jacobian to the eigenvalue zero (cusp singularity). A cusp is therefore defined by the scalar equation
${\vec n}\cdot {\vec e}_0 = 0$.

In order to identify cusps as roots of ${\vec n} \cdot {\vec e}_0$, one needs to find ${\vec e}_0$
as a smooth function along the critical curve. 
However, based on just the local Jacobian
\begin{equation}
\left(\frac{\partial \vec y}{\partial \vec x}\right)
= \left(\begin{array}{cc} y_{11} & y_{12} \\
	 		y_{12} & y_{22} \end{array}\right)\,,
\end{equation}
where $y_{ij} = \partial x_i/\partial x_j$ and $y_{ij} = y_{ji}$, 
there is no such global solution. 
With $y_{11} y_{22} - y_{12}^2 = 0$, one finds that 
\begin{equation}
y_{11} = 0 \vee y_{22} = 0 \Leftrightarrow y_{12} = 0\,,
\end{equation}
i.e.\ the Jacobian is diagonal if and only if one of the diagonal elements
is zero, which reflects the requirement that one eigenvalue has to be
zero. Moreover $y_{11} + y_{22} = 2$, i.e. the trace of the Jacobian
is 2 and the non-zero eigenvalue is 2. Since the condition for a 
critical point implies that $y_{11} y_{22} \geq 0$, one finds $-1 \leq y_{12} \leq 1$, $0 \leq y_{11} \leq 2$, and
$0 \leq y_{22} \leq 2$.
As long as $y_{11} \neq 0$, an eigenvector to the eigenvalue zero is given by
${\vec e}_0^{(1)} = (-y_{12},y_{11})$, but it flips over as $y_{11}$ passes through zero, since
$y_{12}$ changes sign at that instance. Similarly, 
${\vec e}_0^{(2)} = (y_{22},-y_{12})$ is an eigenvector to the eigenvalue zero for $y_{22} \neq 0$, which
flips over as $y_{22}$ passes through zero. This means that the branches of
the critical curves or caustics need further subdivision. 

\begin{figure*}
\includegraphics[width=168mm]{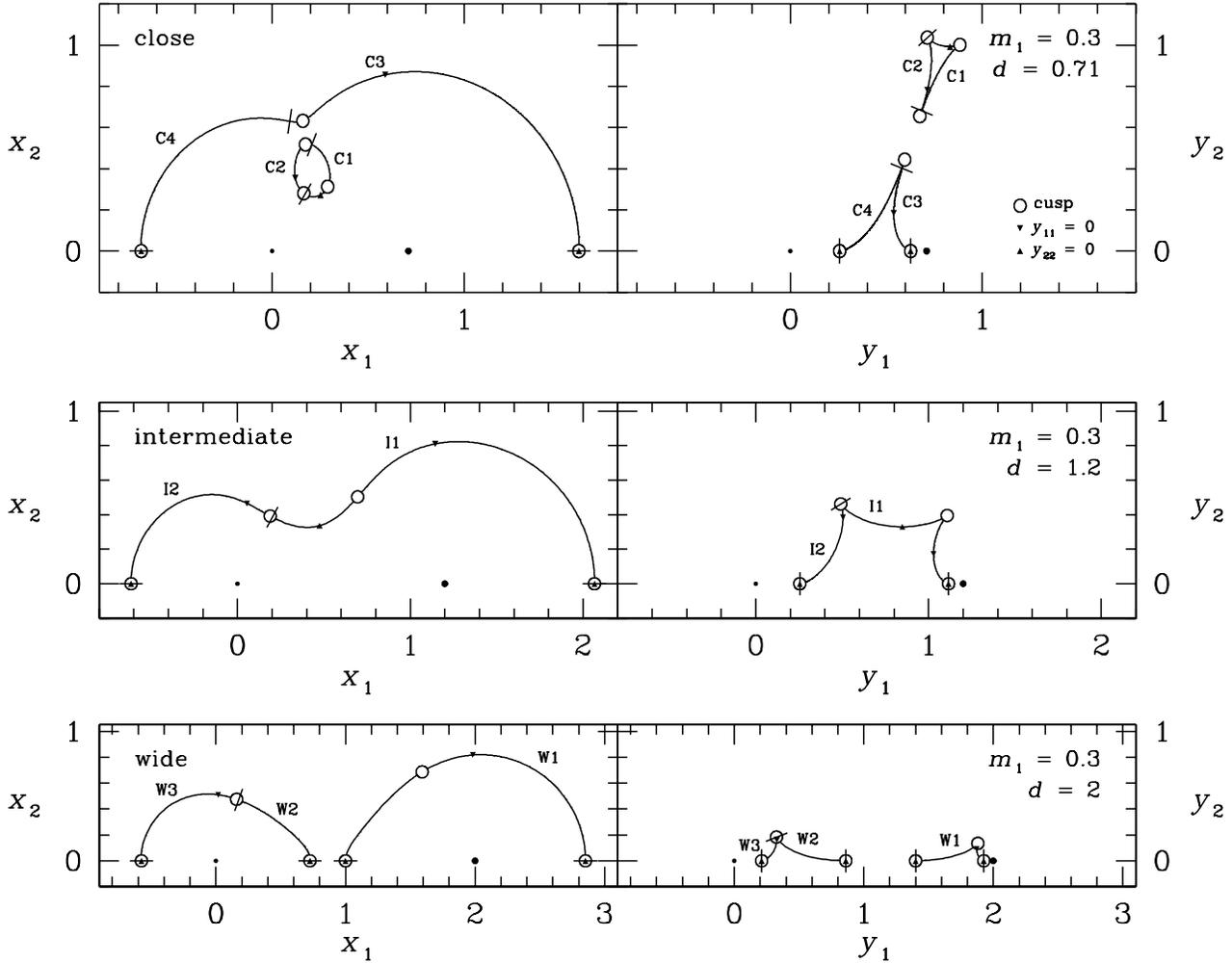}
\caption{Branches of the critical curves and caustics for the three different topologies (wide, intermediate, close). Perpendicular lines separate the different ranges for $r$ and $\cos \varphi$ as
listed in Table~\ref{tab:branches}, whereas triangles mark the points for which
the Jacobian becomes diagonal, i.e.\ either $y_{11} = 0$ (pointing downward)
or $y_{22} = 0$ (pointing upward), and cusps are indicated by circles. Filled dots mark the
position of the two point masses, where the respective areas are proportional to the fractional mass.
The nomenclature of the branches corresponds to Table~\ref{tab:branches}. For $d \leq d_\mathrm{t}$, branch C4 vanishes, while
C3 extends between both intersections of the critical curve with the $x_1$-axis or both intersections of the caustic with
the $y_1$-axis, respectively.}
\label{fig:findcusp}
\end{figure*}

However, the relative positions of the cusps, the points where the Jacobian is diagonal and the points where the branches
of $\cos \varphi$ merge, are unique characteristics for each topology, as
shown in Fig.~\ref{fig:findcusp}. Without restriction of generality, one can assume $m_1 \leq 0.5$
(otherwise simply mirror the coordinates with respect to the vertical axis).
From this figure, one can read off how
to determine the positions of the cusps as well as those points for which
$y_{11} = 0$ or $y_{22} = 0$ by sucessive bracketing using an interval
bisection algorithm for root finding. 
 
Choosing ${\vec e}_0^{(2)}$ as the adaptation of ${\vec e}_0$ for
all $\lambda$ where $y_{22} \neq 0$, minimized the required number of 
subbranches, since several positions with $y_{22} = 0$ coincide with the 
points where the critical curve crosses the $x_1$-axis or the respective
caustic crosses the $y_1$-axis.

\subsection{Finding caustic-associated images}
\label{sec:getcritimg}

If the source includes a cusp, there is no image area that is disconnected from the image
position(s) of the source centre. That case can only arise if the source contour crosses a contiguous fold line between cusps twice. Since the caustic ${\vec y}_\rmn{c}(\lambda)$ is a smooth function of the parameter $\lambda$ between cusps, a circular contour centered around ${\vec y}^{(0)}$ can only cross it twice if the distance
\begin{equation}
d(\lambda; {\vec y}^{(0)}) \equiv \left[{\vec y}_\rmn{c}(\lambda) - \vec y^{(0)}\right]^2
\end{equation}
exhibits a local minimum, so that 
\begin{equation}
\frac{\rmn{d}d(\lambda; {\vec y}^{(0)})}{\rmn{d}\lambda} =
\frac{\rmn{d}{\vec y}_\rmn{c}(\lambda)}{\rmn{d}\lambda} \cdot \left[{\vec y}_\rmn{c}(\lambda) - \vec y^{(0)}\right] = 0
\end{equation}
constitutes a necessary condition,
which means that the tangent to the caustic 
has to be perpendicular to the line connecting it with the source centre.
Only such points can therefore constitute the remaining image positions that
need to be inserted into the adaptive grid in order to ensure that all image
contours are found. 

If the parameter $\lambda$ is chosen as length of the critical curve, one finds
\begin{equation}
\frac{\rmn{d}}{\rmn{d}\lambda} = n_1 \frac{\partial}{\partial x_2} - n_2 \frac{\partial}{\partial x_1}\,,
\label{eq:lambdader}
\end{equation}
with $\vec n$ being is the unit normal vector.

If the curvature $\kappa_{\vec y}(\lambda)$ of the caustic is a monotonic function over the considered range
$[\lambda_\rmn{min},\lambda_\rmn{max}]$, the distance $d(\lambda; {\vec y}^{(0)})$ can only have a single extremum. 
Therefore, for each caustic segment, a potential root of $\rmn{d}\kappa_{\vec y}/\rmn{d}\lambda$ is determined, providing
the final subdivision. With ${\vec t}_{\vec y}(\lambda)$ denoting the caustic tangent, the curvature is given as
\begin{equation}
\kappa_{\vec y}(\lambda) = \frac{t_{\vec y,1} t'_{\vec y,2} - t'_{\vec y,1} t_{\vec y,2}}
{\left|{\vec t}_{\vec y}\right|^3}\,,
\end{equation}
where the prime denotes the derivative with respect to $\lambda$, as defined by Eq.~(\ref{eq:lambdader}).
By further differentiation with respect to $\lambda$, $\rmn{d}\kappa_{\vec y}/\rmn{d}\lambda$ arises
as analytical function of up to the 4th derivatives of ${\vec y}(\vec x)$ with respect to $x_1$ or $x_2$.

\section{Applications and limitations}
\label{sec:lightcurves}

\begin{figure*}
\includegraphics[width=168mm]{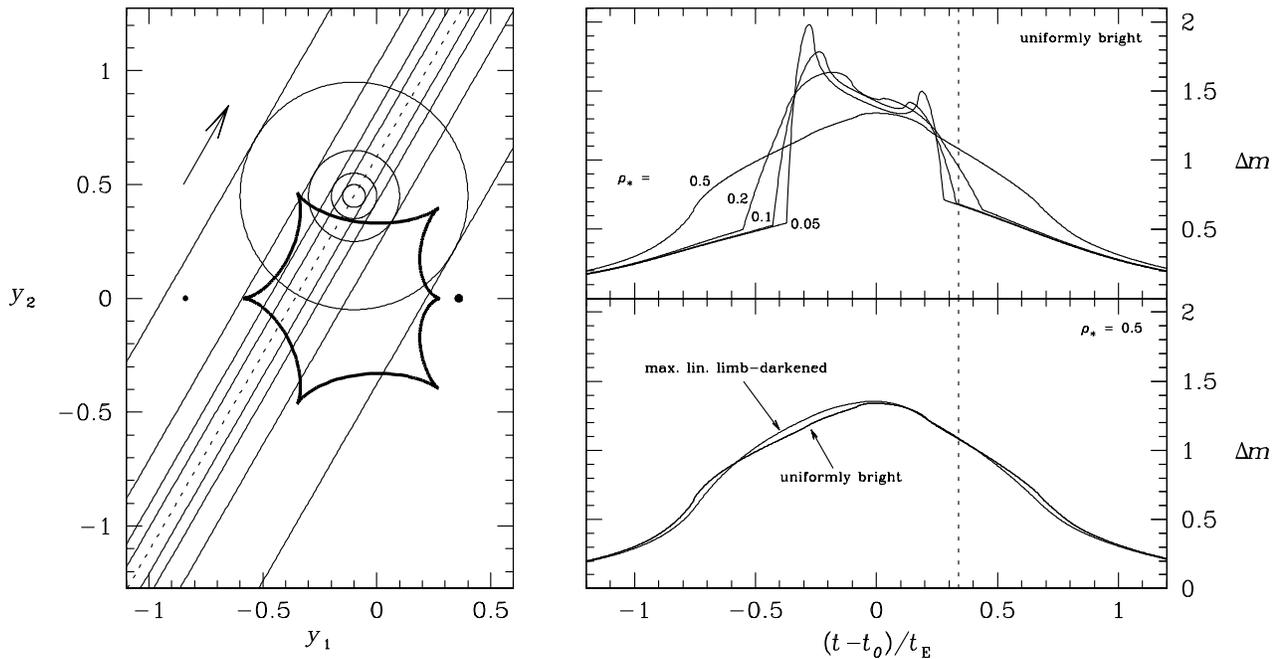}
\caption{Example light curves for sources of different radii and with different
brightness profiles affected by a binary lens. The components of the lens
binary have mass fractions
$m_1 = 0.3$ (left) and $1-m_1 = 0.7$ (right) and their separation
in units of the angular Einstein radius $\theta_\rmn{E}$ is given
by $d = 1.2$. The source is assumed to move uniformly with
respect to the lens at an angle of $60^{\circ}$ with respect to the
binary-lens axis and to pass through $\vec y^{(0)} = (-0.1,0.45)$, where
the coordinates $\vec y$ refer to the angular source position in units
of the angular Einstein radius $\theta_\rmn{E}$ with respect to the 
centre of mass of the lens system. The arrow shows the direction of motion,
where the source moves by the angle $\theta_\rmn{E}$ within the time $t_\rmn{E}$.
Circles show the snapshot of the source for different radii as it passes through $\vec y^{(0)} = (-0.1,0.45)$,
while straight lines limit the region of the plane traced by it and a dotted line shows the trajectory of the
source centre. A thick line shows the caustic, while the 
positions of the components of the binary lens are indicated by dots whose sizes reflect the mass fractions.
The epoch $t_0$ refers to the closest approach of the source to the centre of mass of the lens binary, whereas
the time at which the source reaches $\vec y^{(0)} = (-0.1,0.45)$ is indicated by a vertical dashed line
in the plots showing the light curves.
The magnification is shown as decrease in magnitude $\Delta m = 2.5 \log A$.}
\label{fig:lightcurves}
\end{figure*}

Fig.~\ref{fig:lightcurves} shows example light curves for circular sources with different radii
and different brightness profiles affected by a binary gravitational lens, characterized
by the separation parameter $d = 1.2$ and the mass fraction $m_1 = 0.3$.  
For the smaller source radii ($\rho_\star = 0.05$ and $\rho_\star = 0.1$),
the source transits the caustic and moves completely inside,
so that separated fold-caustic passages produce distinct characteristic peaks that can
be described by a generic profile function \citep*[e.g.][]{Do:fold}. In these cases,
the insertion of a seed image mapping to a local minimum of the distance of the source
centre from the caustic was essential. In contrast, the source never moves completely
inside the caustic for $\rho_\star = 0.2$ or $\rho_\star = 0.5$. For $\rho_\star = 0.2$,
there are epochs for which the source crosses the same fold line (as for the 
smaller radii) or different adjoining fold lines, with or without the cusp in between.
Moreover, for $\rho_\star = 0.5$ there are epochs for which a various number of up to
three cusps are enclosed by the source.

For $\rho_\star = 0.5$, a light curve for a brightness profile corresponding to
maximal limb darkening, i.e.\
\begin{equation}
\xi(\rho) = \frac{3}{2}\,\sqrt{1-\rho^2}\,,
\end{equation}
is shown along with that for a uniformly bright source. With a smaller fraction
of the total brightness in the outer parts for the limb-darkened source,
the source magnification shows a smaller rise as the source enters or exits the caustic
but is larger in between.

The application of Green's theorem to replace the integration over the
image area by an integration along its boundary is a very efficient approach
if the images are 
moderately distorted, so that a large area for the given boundary length is
enclosed (a circle is optimal). There is significantly less gain, however, for
extremely strong distortions leading to the enclosed area resembling 
a line. This case is indeed approached if a source star gets very
closely aligned with a lens star that is only associated with
much less massive companions such as orbiting planets. Resulting in
large peak magnifications, such configurations are of some specific 
interest due to their planet-detection potential \citep{GS:HME,Ratt:high}.
For modelling the event with the largest peak magnification recorded
so far, OGLE 2004-BLG-343 with $A_0 \sim 3000$, \citet{Dong:ray} have
derived a more efficient variant of the ray-shooting technique.
In fact, the current version of the adaptive contouring algorithm is 
significantly slowed down for very small impact angles between source and 
planet-surrounded lens star, but, for uniformly bright sources, the computation of a single magnification
with a relative uncertainty of $5\times 10^{-4}$ can still be carried out
in $\sim 200$~ms on a 600~MFlops machine 
for magnifications in the range $A \sim 100$--$1000$,
while a result is obtained about 20 times faster ($\sim 10$~ms) for 
moderate $A \sim 3$. It turns out that the same choice for the
accuracy parameter $\varepsilon$ (Sect.~\ref{sec:area}) provides a more accurate result for
larger magnifications. For limb-darkened sources, one needs to add
another numerical integration, which slows down the computation a lot. 
Determining the magnification of limb-darkened sources by integrating along the image boundary therefore becomes inefficient for highly-distorted
images. In this case, alternative techniques should be used. However,
the adaptive grid approach can be combined with other evaluations of
the image area or magnification. Moreover, limb darkening usually adds
a small shift to the magnification, so that all other parameters can be
well approximated by assuming a uniformly bright source during an initial
search (allowing faster computation), whereas limb darkening only needs to
be taken into account in a final refinement step.
For source stars idealized to be uniformly bright, the adaptive contouring
algorithm appears to be much faster than ray-shooting for moderate magnifications, while providing a smaller gain for huge magnifications,
which however can still be
substantial ($\sim 4$) for $A \sim 100$. Since the computation of 
ray-shooting magnification maps for a fixed $(d,q,\rho_\star)$ takes
at least some minutes, the evaluation of a several hundred data points
by adaptive contouring usually appears to be faster.

\section{Summary and conclusions}
\label{sec:summary}
The presented adaptive contouring algorithm provides an efficient way
for calculating microlensing light curves, i.e.\ the combined magnification
of the images as a function of time. While the use of an adaptive grid
prevents wasting time by densely covering regions of little interest,
it can be ensured that for binary lenses no image is missed. This is achieved by finding the images of the source centre by solving
a fifth-order complex polynomial \citep{WM95:fifth}, realizing that holes in the
images must include the position of one of the lens components, and
by identifying a position inside images stretching over a critical curve
making use of its parametrization found by \citet{Erdl}.

The efficiency of the proposed algorithm can be enhanced by using a
prioritization scheme for the subdivision of squares containing the
image contour line (double-adaptive contouring) rather than using a
common resolution depth, by using a better estimate on the uncertainty
of the calculated magnification, or using a better approximation of
the contour line and its enclosed area or the source magnification.

While the performance of the current implementation is impressive for
weakly-distorted images, the computation gets significantly slowed down
if images are strongly distorted, i.e. nearly degenerate into
lines. For limb-darkened sources, the evaluation of the magnification by
an integration along a contour line becomes inefficient for such images. However, the adaptive grid approach could be combined with other evaluation algorithms.

As already pointed out by \citet{Do98:NumSrc}, the presented approach
can also be used for calculating the astrometric microlensing signal,
i.e.\ the positional variation of the centroid of light composed of
the images.

The identification of cusps and the parametrization of the branches of
the critical curve in between can also serve as basis for algorithms
searching for matching binary-lens parameters of microlensing features
where the source passes over a fold- or cusp caustic.

An implementation in C is available from the author and can be 
downloaded from {\tt http://star-www.st-and.ac.uk/\~{}md35/Software.html}.

\section*{Acknowledgments}
I would like to thank Keith Horne for reading and commenting on the
manuscript, and acknowledge support by PPARC rolling grant
PPA/G/O/2001/00475. Special thanks go to Subo Dong for running some tests
with his modified ray-shooting code that allowed a comparison of 
efficiency.

\bibliographystyle{mn2e}
\bibliography{adaptive}

\begin{thebibliography}{}

\bibitem[\protect\citeauthoryear{{Dominik}}{{Dominik}}{1995}]{Do95:Num}
{Dominik} M.,  1995, A\&AS, 109, 597

\bibitem[\protect\citeauthoryear{{Dominik}}{{Dominik}}{1998}]{Do98:NumSrc}
{Dominik} M.,  1998, A\&A, 333, L79

\bibitem[\protect\citeauthoryear{{Dominik}}{{Dominik}}{2004}]{Do:fold}
{Dominik} M.,  2004, MNRAS, 353, 69

\bibitem[\protect\citeauthoryear{{Dong} et~al.,}{{Dong}
  et~al.}{2006}]{Dong:ray}
{Dong} S.,  et~al., 2006, ApJ, 642, 842

\bibitem[\protect\citeauthoryear{{Einstein}}{{Einstein}}{1936}]{Ein36}
{Einstein} A.,  1936, Science, 84, 506

\bibitem[\protect\citeauthoryear{{Erdl} \& {Schneider}}{{Erdl} \&
  {Schneider}}{1993}]{Erdl}
{Erdl} H.,  {Schneider} P.,  1993, A\&A, 268, 453

\bibitem[\protect\citeauthoryear{{Gould}}{{Gould}}{1994}]{Gouldapprox}
{Gould} A.,  1994, ApJ, 421, L71

\bibitem[\protect\citeauthoryear{{Griest} \& {Safizadeh}}{{Griest} \&
  {Safizadeh}}{1998}]{GS:HME}
{Griest} K.,  {Safizadeh} N.,  1998, ApJ, 500, 37

\bibitem[\protect\citeauthoryear{{Kayser}, {Refsdal} \& {Stabell}}{{Kayser}
  et~al.}{1986}]{KRS:rayshooting}
{Kayser} R.,  {Refsdal} S.,    {Stabell} R.,  1986, A\&A, 166, 36

\bibitem[\protect\citeauthoryear{{Rattenbury}, {Bond}, {Skuljan} \&
  {Yock}}{{Rattenbury} et~al.}{2002}]{Ratt:high}
{Rattenbury} N.~J.,  {Bond} I.~A.,  {Skuljan} J.,    {Yock} P.~C.~M.,  2002,
  MNRAS, 335, 159

\bibitem[\protect\citeauthoryear{{Schneider} \& {Wei{\ss}}}{{Schneider} \&
  {Wei{\ss}}}{1986}]{SchneiWei:twomass}
{Schneider} P.,  {Wei{\ss}} A.,  1986, A\&A, 164, 237

\bibitem[\protect\citeauthoryear{{Schneider} \& {Wei{\ss}}}{{Schneider} \&
  {Wei{\ss}}}{1987}]{SchneiWei:AGN}
{Schneider} P.,  {Wei{\ss}} A.,  1987, A\&A, 171, 49

\bibitem[\protect\citeauthoryear{{Schneider}}{{Schneider}}{1985}]{Schneider:th%
eory}
{Schneider} P.,  1985, A\&A, 143, 413

\bibitem[\protect\citeauthoryear{{Schramm} \& {Kayser}}{{Schramm} \&
  {Kayser}}{1987}]{SK87}
{Schramm} T.,  {Kayser} R.,  1987, A\&A, 174, 361

\bibitem[\protect\citeauthoryear{{Witt} \& {Mao}}{{Witt} \& {Mao}}{1994}]{WM94}
{Witt} H.~J.,  {Mao} S.,  1994, ApJ, 430, 505

\bibitem[\protect\citeauthoryear{{Witt} \& {Mao}}{{Witt} \&
  {Mao}}{1995}]{WM95:fifth}
{Witt} H.~J.,  {Mao} S.,  1995, ApJ, 447, L105

\end{thebibliography}

\end{document}